\documentclass{camera}
\usepackage{graphicx}  

\begin{document}

%
\title{Ultra high energy neutrinos: the key to ultra high energy cosmic rays}

%
\author{Todor Stanev}

%
\organization{Bartol Research Institute, Department of Physics
 and Astronomy, University of Delaware, Newark, DE~19716, U.S.A. }

\maketitle

\begin{abstract}
 We discuss the relation between the acceleration spectra of
 extragalactic cosmic ray protons and the luminosity and cosmological
 evolution of their sources and the production of ultra high
 energy cosmogenic neutrinos in their propagation from the
 sources to us.  
\end{abstract}
\vspace{1truecm}


\section{Introduction}
  High energy astrophysical neutrinos are produced in hadronic
 reactions of accelerated nucleons and the subsequent decay 
 of the secondary meson and muons. In astrophysical environments
 all mesons and muons decay, so that the generated neutrino
 flux indicates the energy spectrum of the accelerated nucleons
 and the available target density in the neutrino production site,
 as it does for $\pi^0$ decay $\gamma$-rays.
 One of the best examples for this correlation is the prediction
 for diffuse $\gamma$-ray fluxes from the galactic plane.
 To fit the observations the EGRET group had to use the data 
 on the matter density and the cosmic ray density in the galactic
 plane (Hunter et al. 1997).

  Waxman\&Bahcall (1999, 2001) did the same kind of calculation
 of the  high energy neutrinos generated in extragalactic
 sources replacing the target density with a parameter $\eta$
 that describes the fraction of the accelerated cosmic rays at 
 all extragalactic sources that interact at source and 
 generate neutrinos. The process is photoproduction 
 interactions of the accelerated protons in the ambient photon
 field. One very important parameter is the cosmic ray
 emissivity that was estimated (Waxman 1995) to be 
 4.5$\times$10$^{44}$ erg/Mpc$^3$/yr with an error of about
 30\%. The acceleration proton spectrum is assumed to be
 $E_p^{-2}$ and the maximum proton energy is assumed to be
 10$^{21}$ eV. The emissivity above refers to the 
 cosmic ray flux measured at 10$^{19}$ eV, which is assumed to 
 be of extragalactic origin. The calculation uses the average
 fraction of the proton energy that neutrinos carry - 15\% in
 all neutrino flavors. The point made is that if $\eta$ = 1,
 i.e. if all protons interact, this will give an upper bound
 of the astrophysical neutrino flux.

  This bound was criticized by Mannheim, Protheroe \& Rachen (2001)
 where a more
 detailed analysis of the assumptions made by Waxman\&Bahcall (1999)
 was performed and a new, generally higher, limit that accounts
 for the energy loss horizon of cosmic rays and neutrinos was
 derived. The work of Waxman\&Bahcall, right or wrong, is the first 
 {\em published} direct connection of the flux of astrophysical
 neutrinos and of extragalactic cosmic rays.     

  We shall attempt to relate the flux of ultra high energy cosmic
 rays (UHECR) to UHE neutrinos created in the propagation 
 of these particles from their sources to us in the photon
 fields present in the whole Universe. The main one is the
 microwave background radiation (MBR) where most of these
 {\em cosmogenic} neutrinos are generated. Neutrino production
 in the GZK interactions was first proposed by Berezinsky\&Zatsepin
 (1969). Many other calculations were subsequently 
 made and the most important of those is that of Hill\&Schramm (1983)
 who attempted to limit the cosmological evolution of the 
 cosmic ray sources by the lack of detection of such ultra
 high energy neutrinos.

  The difference between the source neutrinos discussed by
 Waxman\&Bahcall and the cosmogenic neutrinos is that the target
 for the latter is extremely well studied. While different
 astrophysical sources show various photon spectra the MBR 
 energy spectrum is not only well known in the current epoch,
 but can be easily evolved for any value of the redshift $z$.
 Because of that we know exactly what the threshold proton 
 energy $E^{thr}_p$ for photoproduction is:
 $$E^{thr}_p \, = \, \frac{m_\pi (2m_p + m_\pi)}
 {2 \varepsilon (1 - \cos \theta)}\; ,$$
 where $\varepsilon$ is the photon energy and $\theta$ is 
 the angle between the two interacting particles.
 $E^{thr}_p$ in the current epoch is above 2$\times$10$^{20}$ eV
 for interactions on the average energy MBR photon. The
 actual threshold is 
 about 3$\times$10$^{19}$ eV. For earlier cosmological epoch 
 it decreases as $(1 + z)^{-1}$ because of the increase of the
 MBR temperature. In the following we shall assume that all
 extragalactic cosmic rays are protons and that their sources
 are isotropically and homogeneously distributed in the Universe.

  The mean free path for proton photoproduction interactions 
 in the contemporary Universe reaches a minimum of 3.8 Mpc
 (Stanev et al 2000)
 at proton energy of 5$\times$10$^{20}$ eV and very slightly
 increases after that. The energy loss length for protons
 is 16 Mpc at the same energy (the inelasticity coefficient $K_{inel}$
 for photoproduction interactions at threshold is below 0.2)
 and decrease at higher energy when $K_{inel}$ continues growing
 with $\sqrt{s}$. 

  To obtain the flux of cosmogenic neutrinos we calculate the 
 neutrino production yield at $z$ = 0 and then scale it for
 (Engel, Seckel \& Stanev 2001) arbitrary redshift values.
 The flux is obtained by folding
 the neutrino yield with the proton injection spectrum and
 integration in time or redshift in a cosmological model.
 The $z$ dependence of the proton injection spectrum can 
 include the cosmological evolution of its sources which in 
 the current work is assumed to be of the form $(1 + z)^m$
 with different $m$ values in various
 cosmological epochs. We use $H_0$ of 75 km/s/Mpc.
 $\Omega_\Lambda$ dominated Universe enhances the cosmogenic
 neutrino fluxes by about 75\% in comparison with the matter 
 dominated Universe ($\Omega_M$ = 1).  

\section{Fits of the UHE cosmic ray spectrum}

\begin{figure}
\includegraphics[width=5truein]{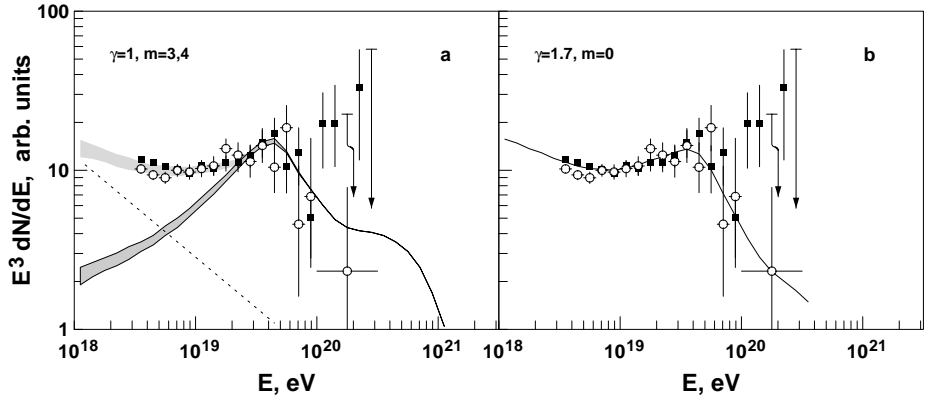}
\caption{Left hand panel: Fit of the observed cosmic ray spectrum with 
 flat injection spectrum (\protect$\gamma=1$) and cosmological evolutions
 of the cosmic ray sources with $m$ = 3 and 4 (bottom and top of the
 shaded area). The galactic cosmic ray spectrum is shown with a dashed line.
 Right hand panel: Fit with steep injection spectrum (\protect$\gamma=1.7$)
 and no cosmological evolution.
\label{spectrum}}
\end{figure}

 One of the current problems in ultra high energy astrophysics
 is the derivation of the extragalactic UHECR acceleration spectrum
 which for the purposes of propagation studies we also call {\em injection} 
 spectrum. Fig.~\ref{spectrum} shows two extreme fits. The left hand one 
 is a fit with $E_p^{-2}$ extragalactic injection spectrum
 (Bahcall\&Waxman 2003) and two different 
 cosmological evolutions with $m$=3 and 4 up to a $z_{max}$ = 1.9.
 The small difference between $m$ = 3 and 4 is explained with the
 fact that only readshifts smaller than 0.4 contribute to the 
 UHECR flux above 10$^{19}$ eV. Since the contribution of extragalactic
 protons below 10$^{19}$ eV is small one has to assume that the
 galactic cosmic ray spectrum extends up to and above that energy.
 The experimental points shown are these of AGASA
 (Takeda et al 1998) and HiRes
 (Abbasi et al 2004) normalized to each other at 10$^{19}$ eV to 
 emphasize the shape of the UHECR spectrum. Except for the AGASA
 events above 10$^{20}$ eV the two spectra agree quite well in shape.
 The preliminary data of the Auger Observatory (Sommers 2005) have
 a flux normalization close to that of HiRes and a similar energy
 spectrum.
 
 The right hand figure shows a different fit
 (Berezinsky, Gazizov \& Grigorieva 2005) -
 the extragalactic injection spectrum is $E_p^{-2.7}$ and there is
 no  cosmological evolution of the UHECR sources (m = 0).
 The fit describes quite well the cosmic ray spectrum down to
 at least 10$^{18}$ eV. The wide feature in the spectrum
 around 10$^{19}$ eV is explained with the second most important
 energy loss process in proton propagation - the Bethe-Heitler
 production of $e^+e^-$ pairs (Berezinsky \& Grigorieva 1988).
 According to this fit
 the end of the galactic cosmic rays is at 10$^{18}$ eV or below.

 In addition to the different ends of the galactic cosmic ray spectrum
 (and respectively the cosmic ray chemical composition as a function
 of the energy)
 the major difference between the two fits is the cosmological
 evolution of the cosmic ray sources that they require. Fits with 
 flat injection spectrum ($E_p^{-2}$, $\gamma$ = 1) require a strong
 cosmological evolution of the cosmic ray sources, similar to that
 of star forming regions. The fit of Berezinsky et al (2005)
 on the other hand does not need any cosmological evolution of the
 sources. Our own fits have shown that if cosmological evolution
 is assumed the best value for the injection spectrum index $\gamma$ 
 decreases by 0.05 to 0.15 (DeMarco \& Stanev, 2005).

\section{Redshift dependence of the cosmogenic neutrino production}

 Since neutrinos do not lose energy in propagation their production
 follows well the cosmological evolution of the cosmic ray sources.
 To illustrate that point we briefly discuss calculation of the
 cosmogenic neutrino flux. The 
 neutrino flux at Earth due to GZK process is
\begin{equation} 
\label{ss:eq1}
E_\nu \frac{d\Phi}{dE_\nu}(E_\nu) = \int dt d\epsilon_p 
\frac{d\Gamma}{d\epsilon_p} 
E_\nu \frac{dy}{dE_\nu}(E_\nu,\epsilon_p,t) 
\end{equation}
 where $\Gamma$ is the injection rate of UHECR and $y$ is the neutrino yield 
 per proton injected with energy $\epsilon_p$, and $E_\nu$ is the neutrino 
 energy today. Equation \ref{ss:eq1} can be put into a more convenient
 form by defining $q=1+z$ and integrating over redshift. For simplicity 
 in this example we assume matter dominated Universe.

 After scaling the neutrino yield from photoproduction interactions in MBR
 as a function of redshift, including a cosmological evolution as $q^m$,
 and changing the variable to $\ln q$ the neutrino flux becomes
 (Seckel \& Stanev 2005)
\begin{equation} 
E_\nu \frac{d\Phi}{dE_\nu}(E_\nu) = \frac{3A}{2} 
\int_0^{q_{max}} d(\ln q) q^{(m+\gamma-\frac{3}{2})} 
E_\nu\frac{dY_{0\gamma}}{dE_\nu}(q^2 E_\nu)
\end{equation}

\begin{figure}
\includegraphics[width=5truein]{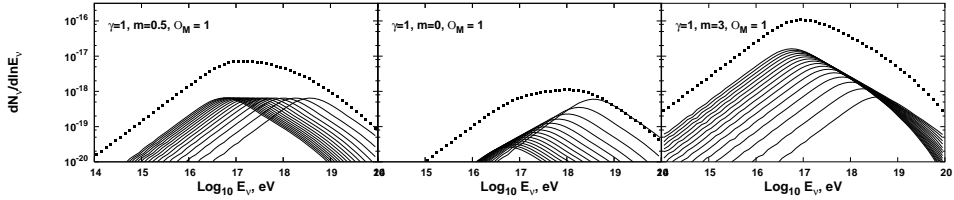}
\caption{Three examples with different injection spectrum and 
 cosmological evolution of the cosmic ray sources. Note that 
 for clarity the examples are made with \protect$\Omega_M$ = 1
 and cosmological evolution extending to redshift of 9. The contributions 
 of different redshifts are plotted at \protect$\Delta z$ = 0.5}
\label{ssf2}
\end{figure}

  It is now clear that for values of $(m + \gamma)$ greater than 3/2 
 higher redshifts generate more neutrinos, while for values below that
 the neutrino production diminishes with redshift. Fig.~\ref{ssf2}
 shows three illustrative examples where the same cosmological evolution
 is assumed to $q$ = 10.  

 In understanding the importance of Fig.~\ref{ssf2} one should 
 remember that redshifts above 0.4 do not contribute to the 
 extragalactic cosmic ray spectrum above 10$^{19}$ eV independently
 of the maximum acceleration energy. For this reason the cosmological
 evolution of the sources affects the observed cosmic ray spectrum only
 slightly, as visible in Fig.~\ref{spectrum}.

 \section{Cosmogenic neutrinos from the two extreme fits of 
 the UHECR spectrum.}

\begin{figure}
\begin{center}
\includegraphics[width=3.5truein]{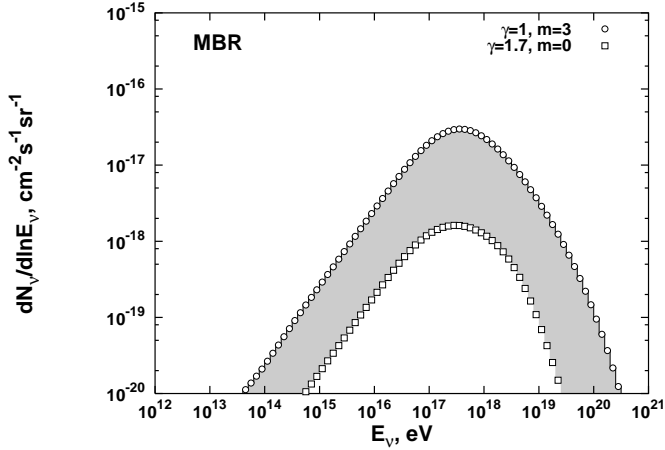}
\caption{Sum of muon neutrinos and antineutrinos generated by the two
extreme fits of the UHECR spectrum
 \label{mbr}}
\end{center}
\end{figure}

 It is now obvious that the two fits of the cosmic ray spectrum
 will generate  different fluxes of cosmogenic neutrinos.
 In the flat injection spectrum case we have $m + \gamma$ = 4,
 which provides for a strong  cosmological evolution of the
 neutrino production. In the steep injection spectrum fit we
 have practically no cosmological evolution. Fig.~\ref{mbr} shows 
 the spectra from the both fits assuming the same cosmic ray flus at
 10$^{19}$ eV, which corresponds to the emissivity derived by
 Waxman (1995).
 The cosmological evolution for the flat injection spectrum model
 ${\cal H} (z)$ used is from the same paper and is
\begin{equation}
  {\cal H}(z) =  \left\{
\begin{array}{lcl}
 (1 + z )^3  &  {\rm :}   &  z <  1.9\\
 (1 + 1.9)^3 &  {\rm :}   & 1.9 < z < 2.7\\
 (1 + 1.9)^3 \exp\{(2.7-z)/2.7\} & {\rm :} &
  z >  2.7
\end{array}\right. 
\label{evo} 
\end {equation}
  The two models indeed generate very different cosmogenic neutrino 
 spectra, which are shown in Fig.~\ref{mbr} for dark energy dominated
 cosmology and $H_0$ = 75 km/s/Mpc. The flat injection spectrum
 model generates about 1.5 orders of magnitude more neutrinos in the
 peak that is at about 10$^{17.6}$ eV. The steep spectrum contains
 a smaller faction of above threshold protons and the neutrino flux
 peaks at slightly lower energy. This will influence also the
 detection probability as
 the neutrino-nucleon cross section grows with the neutrino energy.

 The flat injection spectrum model generates about 10$^{-19}$ muon
 neutrinos and antineutrinos per cm$^2$.s.srad above 10$^{20}$ eV
 which brings it in the range that could be detected by the 
 Auger observatory (Abraham et al 2004) and RICE (Kravchenko et al. 2006)
 and ANITA-like (Barwick et al 2006)
 radio detection experiment. The steep injection spectrum model
 with no cosmological evolution goes below 10$^{-20}$ per cm$^2$.s.ster
 shortly above 10$^{19}$ eV. 

 In practice this means that the cosmogenic neutrinos from the 
 flat injection spectrum model could be detected by the neutrino 
 telescopes in construction, including IceCube (Ahrens et al 2004)
 while the steep injection spectrum model would not allow us 
 to detect cosmogenic neutrinos in the foreseeable future.

 It is indeed very difficult to estimate the detection rate in
 different experimental arrangements without the use of a proper
 detection montecarlo code that can estimate the detection 
 probability as a function of the neutrino energy and 
 direction. Roughly speaking, the IceCube neutrino telescope should 
 be able to see about one event per year with energy above 10$^6$ 
 GeV from the flat injection spectrum model. The large majority of
 these events will come from above, as most neutrinos coming from
 the lower hemisphere will be absorbed in propagation through the Earth.
 The cosmogenic neutrinos generated by the steep injection spectrum
 model will generate detection rates smaller by more than an order
 of magnitude.

 The possible future detection even of a very small number of
 neutrino events of extremely high energy could serve as a
 powerful tool for solving the problem of the acceleration spectrum
 of UHECR and the cosmological evolution of their sources.

\section{Cosmogenic neutrinos from proton interactions in other 
 universal photon fields}

 The microwave background radiation is by far not the only 
 universal photon field - it is only the best known. For this
 reason we performed a calculation (DeMarco et al 2006) of the
 neutrino
 production on the infrared and optical background (IRB). 
 This background has been measured several times, directly and
 indirectly by the TeV gamma ray absorption, but its exact 
 number density is a matter of dispute. We use the model of 
 Stecker, Malkan\& Sculy (2005) which also 
 follows the cosmological evolution of the infrared background.
 We are mostly interested in the far infrared background (FIR) 
 that has the highest number density and provides a denser target
 for high energy proton interactions. 

\begin{figure}[h]
\begin{center}
\includegraphics[width=3.5truein]{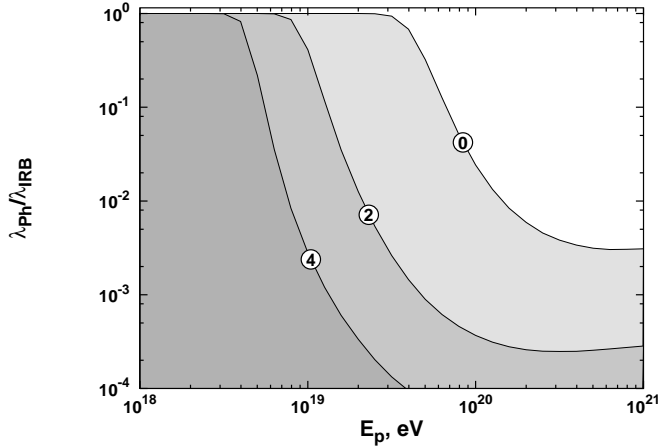}
\caption{Ratio of the proton mean free path in the total photon
 field to this in the infrared background. 
 \label{irbg}}
\end{center}
\end{figure}

 Figure~\ref{irbg} shows the ratio of the proton interaction length 
 in the ratio of the proton mean free path in all photon fields with
 photon energy below 1 eV to that in the infrared background as a 
 function of the proton energy. There are three curves corresponding
 to redshifts of 0, 2, and 4, i.e. covering almost the whole redshift
 range important for cosmogenic neutrino production. Since the proton
 threshold energy for interactions in MBR at $z$ = 0 is about
 3$\times$10$^{19}$ eV at lower energy all interactions are in the IRB.
 One question is how IRB with number density of less than 1 cm$^{-3}$
 can compete with the MBR with number density of 430 cm$^{-3}$ above
 the threshold energy. The answer is simply that only a small fraction
 of the MBR photons provide above threshold targets for photoproduction
 while almost all IRB photons do. 

 At higher redshifts the dominance of the IRB photon extends to lower
 energy. The main reason is the different cosmological evolution of 
 the two backgrounds. MBR photons energy grows as $(1+z)$ and their
 number density as $(1+z)^3$ while IRB photons have slower 
 cosmological evolution - they are emitted by astrophysical objects
 and we suspect that the IRB number density at redshifts higher than 6 
 is close to zero.
 
\begin{figure}[h]
\begin{center}
\includegraphics[width=3.5truein]{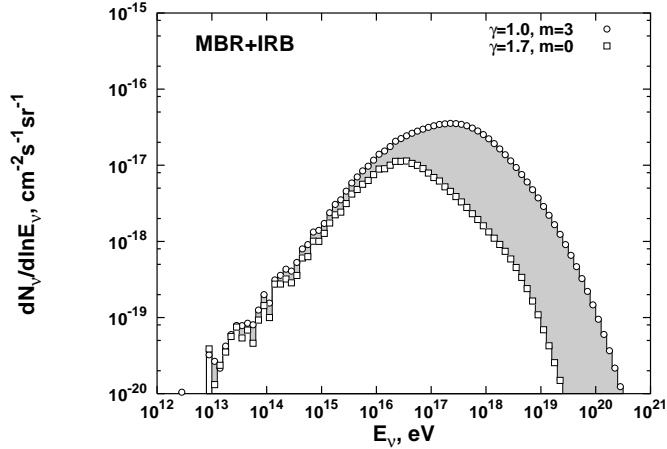}
\caption{Sum of muon neutrinos and antineutrinos generated by the two
 fits of the extragalactic cosmic ray spectrum generated in both MBR and
 IRB.
 \label {allnu}}
\end{center}
\end{figure}

 Figure~\ref{allnu} shows the cosmogenic neutrino fluxes generated
 by the two fits of the UHECR spectrum with account for the 
 interactions in the infrared background. The difference between
 the two models now is much smaller. Since the UHECR emissivity
 is normalized to the flux at 10$^{19}$ eV the IRB neutrinos 
 in the steep injection spectrum model are much larger fraction of
 those generated in the MBR than those in the flat injection 
 spectrum model. In addition, the strong cosmological evolution of
 the cosmic ray sources in the flat injection spectrum case is not
 enhanced by equally strong cosmological evolution of the photon
 target density.  

 Up to about 10$^{15}$ eV the two spectra are almost identical.
 The steep injection spectrum model generates neutrinos that 
 peak at about 10$^{16.3}$ eV and has a spectrum with relatively
 narrow width. Th flat injection spectrum with cosmological evolution
 peaks as in the MBR only case and has a much wider spectrum.
 The difference in the peak values is now smaller, about a factor
 of 3. The ratio in the detection rates has most likely not 
 changed much because the higher energy neutrinos interact with matter
 with correspondingly higher cross section.

 It is worth noting that we have not calculated the neutrino 
 production by protons of energy below 10$^{18}$ eV and
 photons of energy exceeding 1 eV were neglected. In the
 calculation presented by Allard et al (2006) not only 
 all optical photons, but also UV ones, are included.
 Nucleons of energy lower than 10$^{18}$ eV also interact and
 generate neutrinos. Their lower neutrino yield is weighted 
 by the much higher number of lower energy nucleons and the
 neutrino fluxes have wider distributions and peak at lower
 energy.

\section{Discussion}

 We have shown that the fluxes of cosmogenic neutrinos generated
 by extragalactic protons depend very strongly on the cosmological
 evolution of the UHECR sources. The detection of even a small 
 number of such neutrinos will be an important input in the
 solution of the origin of these particles. Currently the observed 
 UHECR spectrum can be equally well be fitted with vastly 
 different cosmic ray acceleration spectra. Fits with flat ($\gamma$ = 1)
 acceleration spectra require strong cosmological evolution, while
 steep ($\gamma$ = 1.7) do not require any cosmological evolution
 of the sources. The acceleration spectrum could be derived
 if the cosmological evolution of the sources were
 determined by observations of cosmogenic neutrinos.
 
 The traditional way to study the same question is through 
 studies of the chemical composition of these particles.
 Cosmogenic neutrinos give one more parameter that should be
 consistent with the changes of the composition if all 
 extragalactic cosmic rays are light (H and He) nuclei. 
 The steep injection spectrum model predicts changes of the
 cosmic ray composition when approaching 10$^{18}$ eV and
 a very low flux of cosmogenic neutrinos.  In the flat 
 acceleration spectrum model the composition changes continue
 up to and exceeding 10$^{19}$ eV. 

 The calculations and estimates presented here assume that 
 ultra high energy cosmic rays are protons. We have not accounted 
 even for a small He contribution in the flux of UHE particles.
 There are, however, in the literature papers that do similar
 calculations in the assumption that UHECR have the same
 chemical composition as the galactic cosmic rays (Allard 2005).
 Under this assumption the injection spectrum of UHECR comes out
 to be intermediate between the two fits discussed above.
 Cosmogenic neutrinos are also produced (Hooper, Taylor \& Sarkar 2005;
 Ave et al 2005) but they are mostly electron antineutrinos 
 from the decay of neutrons from nuclear photodisintegration.
 These neutrinos are also strongly correlated with the cosmological
 evolution of the UHECR sources. 
 
\section{Acknowledgments}

 Parts of the work reported here are done in a close collaboration 
 with D.De~Marco, D.~Seckel and others. This work is supported in
 part by NASA grant ATP03-0000-0080.

\bigskip
\bigskip
\noindent {\bf DISCUSSION}

\bigskip
\noindent {\bf FRANCESCO VISSANI:} You estimated the number of neutrino
events in IceCube to 1.5 per year. What is the expected number of 
events in ANITA and Auger ? 

\bigskip
\noindent {\bf TODOR STANEV:} In principle Auger has 30 times more volume
 than IceCube, but the number of events is probably smaller. I would guess
 it does not exceed 5 events per year depending on the exact energy
 threshold and efficiency. Even for IceCube I gave you my own estimate.


\begin{thebibliography}{9}
\bibitem{}Abbasi, R.U. et al, 2004, Phys. Rev. Lett., 92:151101

\bibitem{}Abraham, J. et al, 2004, (Pierre Auger Collaboration)
 Nucl.Instrum.Meth. A523:50-95

\bibitem{}Ahrens, J. et al, 2004, (IceCube Collaboration)
 Astropart.Phys.20:507-532

\bibitem{}Allard, D. et al, 2005, A\&A, 443:29

\bibitem{}Allard, D. et al, 2006, astro-ph/0605327

\bibitem{}Ave, M. et al, 2005, Astropart. Phys. 23:19

\bibitem{}Bahcall, J.N. \& Waxman, E, 2001, Phys.~Rev.~D64:023002

\bibitem{}Bahcall, J.N. \& Waxman, E., 2003, Phys.Lett.B556:1-6

\bibitem{}Barwick, S.W. et al., 2006 (ANITA Collaboration) 
Phys.Rev.Lett.96:171101

\bibitem{}Berezinsky, V., Gazizov, A.Z. \& Grigorieva, S.I., 2005, 
Phys.Lett.B612:147-153

\bibitem{}Berezinsky, V.S. \& Grigorieva, S.I., 1988, A\&A, 199:1

\bibitem{}Berezinsky, V.S. \& Zatsepin, G.T., 1969, Phys. Lett., 29B:423

\bibitem{}De Marco, D. \&Stanev, T., 2005, Phys.Rev.D72:081301

\bibitem{}De Marco, D. et al, 2006, Phys.Rev.D73:043003

\bibitem{}Engel, R, Seckel, D. \& Stanev, T., 2001, Phys.~Rev.~D, 64:093010

\bibitem{}Hill, T.C. \& Schramm, D.N., 1985, Phys.~Rev.~D31:564 

\bibitem{}Hooper, D., Taylor, A \& Sarkar, S., 2005, Astropart. Phys. 29:11

\bibitem{}Hunter, S.D. et al, 1997, Ap.~J.481:205-240

\bibitem{}Kravchenko, I. et al, 2006, Phys. Rev. D73:082002

\bibitem{}Mannheim, K., Protheroe, R.J. \& Rachen, J.P., 2001, 
Phys.~Rev.~D63:023003

\bibitem{}Seckel, D., \& Stanev, T., 2005, Phys. Rev. Lett., 95:141101

\bibitem{} Sommers, P. (Auger Collaboration) 2005, Proc. 29th International
 Cosmic Ray Conference (Pune, India) 7:387

\bibitem{}Stanev, T. et al, 2000, Phys. Rev. D, 62:093005

\bibitem{}Stecker, F.W., Malkan, M.A. \& Scully, S.T., 2005, astro-ph/0510449

\bibitem{}Takeda, M. et al, 1998, Phys. Rev. Lett., 81:1163

\bibitem{}Waxman, E \& Bahcall, J.N., 1999, Phys.~Rev.~D59:023002

\bibitem{}Waxman, E., 1995, Ap.~J., 452:L1-L4

\end{thebibliography}
\end{document}